\documentclass[showpacs,floatfix,eqsecnum,aps,nofootinbib]{revtex4}
\usepackage{epsfig}

\begin{document}

\title{Spheroidal galactic halos and mirror dark matter}
\author{R. Foot}\email{foot@physics.unimelb.edu.au}
\affiliation{School of Physics, Research Centre for High Energy
Physics, The University of Melbourne, Victoria 3010, Australia}
\author{R. R. Volkas}\email{r.volkas@physics.unimelb.edu.au}
\affiliation{School of Physics, Research Centre for High Energy
Physics, The University of Melbourne, Victoria 3010, Australia}

\begin{abstract}
Mirror matter has been proposed as a dark matter candidate. It has
several very attractive features, including automatic 
stability and darkness, the ability to mimic the
broad features of cold dark matter while in the linear density
perturbation regime, and consistency with all direct dark matter
search experiments, both negative (e.g. CDMS II) and positive (DAMA).
In this paper we consider an important unsolved problem:
Are there plausible reasons to explain why most of the mirror matter 
in spiral galaxies
exists in the form of gaseous {\it spheroidal} galactic halos 
around ordinary matter {\it disks}? We compute an order-of-magnitude 
estimate that the mirror photon luminosity
of a typical spiral galaxy today is 
around $10^{44}$ erg/s. Interestingly, this 
rate of energy loss is similar to the power supplied by ordinary supernova
explosions. We discuss circumstances under which supernova power
can be used to heat the gaseous part of the mirror matter halo and hence prevent
its collapse to a disk. The {\it macro}scopic ordinary-mirror asymmetry plays
a fundamental role in our analysis.

\end{abstract}


\maketitle

\section{Introduction}

Mirror matter, a hypothetical parity-transformed partner for
ordinary matter, is a simple and well-defined extension\cite{flv} 
of the standard
model of particle physics that has interesting cosmological
consequences. In particular, since the microphysics within
the mirror sector is identical to that of the ordinary 
sector\footnote{Except that mirror weak interactions are right-handed
while ordinary weak interactions are left-handed, a distinction
that will not be important for the physics discussed in this paper.},
mirror electrons $e'$ and mirror protons $p'$ will be as long-lived
as ordinary electrons and protons (we will denote mirror partners of
ordinary particles by a prime). Mirror atoms or ions thus pass
the first important test for dark matter candidature: if they
were created in the early universe, they will still exist today.
Another important test -- very weak coupling to ordinary photons --
follows naturally if the ordinary and mirror sectors are in general almost
decoupled in all senses except the gravitational, a circumstance
that is easily arranged (see below). Further, mirror dark matter
behaves similarly to, but not identically with, cold dark matter at galactic 
and larger scales during the linear regime of density perturbation
growth\cite{comelli,rays}. 
Moving partially to the terrestrial domain, {\it all} existing 
direct dark matter detection experiments are consistent with
the mirror dark matter hypothesis, whether those experiments have
yielded negative results (e.g. CDMS II\cite{cdms2}, Edelweiss\cite{ed}, etc.) or
positive (DAMA)\cite{dama}, as has been explained at length by one of us
recently\cite{footdama}. Mirror matter also has other desirable properties,
as reviewed in Ref.\cite{review}.

But mirror matter has a potential Achilles heel. The purpose
of this paper is to discuss aspects of this important
problem, and to suggest possible solutions. The problem
is familiar to astrophysicists:
For mirror matter to be an important dark matter component,
its distribution in spiral galaxies such
as the Milky Way must be spheroidal.
Ordinary matter,
on the other hand, has collapsed into the bulge and disk.
These different macroscopic behaviours of ordinary and 
mirror matter demand an explanation.

Mirror matter can exist in compact form (mirror stars, planets, and so on) and
as a gas component. Evidence for compact halo objects  
has emerged from microlensing surveys of the Large Magellanic Cloud\cite{macho} and M31\cite{new}.
These observations are 
consistent with a halo containing a mass fraction of $f \sim 0.2$ 
in the form of mirror stars\cite{machoexp}, with a $95\%$ confidence
interval of $0.08 < f < 0.50$\cite{macho}.
This suggests that a significant
gas component $1-f \sim 0.8$ should exist in the spheroidal halo.\footnote{This
is true if mirror matter comprises the entire dark matter sector,
the hypothesis we adopt here. It is of course possible that
mirror matter contributes to but does not exhaust the
dark matter sector. For instance, spheroidally distributed mirror matter
compact objects at the level of 
$8-50\%$ of the halo could explain the MACHO observations, with
the diffuse dark component ascribed to something else (axions, etc.).
Mirror matter gas would then be free to exist in cooled
form in the disk, provided it is not overabundant. It is the
roughly spherical distribution of the mirror {\it gas} component that
is the main problem. 
One of us (RRV) thanks M. Yu.\ Khlopov for emphasising this to him.} 
Evidence for a significant gas
component also arises from the DAMA annual modulation
signal\cite{dama,footdama}. The gas component must be supported
against gravitational collapse via its pressure
(being spherically distributed, it 
cannot be rotationally supported).
To be consistent, the cooling time scale of the mirror
gas needs to be long -- comparable to the age of the galaxy.


We suspect that the different behaviour of ordinary and 
mirror matter in galaxies is related to 
macroscopic differences between the
ordinary and mirror sectors, which is possible
even if the microphysics is exactly symmetric.
There are three main reasons for this macroscopic mirror
asymmetry.
As has been studied at length (see e.g.\ Ref.\cite{old,comelli}) 
successful big bang nucleosynthesis 
requires the mirror sector temperature during that epoch to
be less than about half the ordinary sector temperature.
Second, to make the mirror Silk damping scale sub-galactic
requires a similar (actually slightly stronger) inequality\cite{comelli,rays}.
Third and most definitively, an impressive body of observational
evidence has established that the ratio of ordinary to
non-baryonic dark matter must be in the range $0.20\pm 0.02$\cite{wmap},
ruling out equal proportions by a comfortable 
margin.\footnote{One way the required cosmological ordinary/mirror asymmetry
can be explained is through inflationary models\cite{inflation,fv}.}
Because of the temperature and density asymmetries, all the
details of chemical abundances and star and galaxy formation
and evolution will be quite different in the two sectors.
For instance, mirror primordial nucleosynthesis will produce
more mirror helium $He'$ than mirror hydrogen $H'$, in contrast
to the ordinary case.
It is possible that the spheroidal-mirror-halo versus
ordinary-disk dichotomy is another instance of the different
initial conditions and subsequent histories.

\section{Estimate of the cooling time}

The purpose of this section is to estimate the cooling time
of the mirror gaseous `halo' in the absence of any significant
heat source.
Estimates of the cooling time of a mirror gaseous halo
can be determined using the standard calculation
for the cooling time of the proto-galactic nebula (see e.g.\ \cite{astro,bin}).
One first establishes using the virial theorem that 
the gas will be fully ionised.
This fact is then used to justify the computation of the energy loss rate and 
hence cooling time of the nebula due to dissipative processes such as 
bremsstrahlung.

Taking the gas to be undergoing quasistatic evolution,
the total kinetic energy $K$ of the gas is
related to the total potential energy $U$ via the virial theorem,
\begin{equation}
-2 K = U.
\end{equation}
The potential energy of a gravitationally bound, spherical  
distribution of constant density is
\begin{eqnarray}
U = -{3 \over 5}{G M^2 \over R},
\end{eqnarray}
where $M$ is the total mass of the nebula and $R$ is its
radius. [Departures from constant density will change the
prefactor $3/5$ to another number of order one. The
constant density idealisation is good enough for the
present purpose.]
Taking a gas of $N$ particles, with
a mean mass of $\mu m_p$ ($m_p$ is the proton mass), 
the virial theorem implies that
\begin{eqnarray} 
-2N{1 \over 2} \mu m_p \langle v^2 \rangle
= -{3 \over 5}{G M^2 \over R},
\label{3}
\end{eqnarray}
where $\langle v^2 \rangle$ is the mean squared-speed
of the gas particles.
The virial temperature of the gas is defined by
\begin{eqnarray}
{1 \over 2} \mu m_p \langle v^2 \rangle = {3 \over 2} kT_{virial},
\end{eqnarray}
which, combined with
Eq.(\ref{3}), yields
\begin{eqnarray}
kT_{virial} = {\mu m_p G M \over 5R}.
\end{eqnarray}
Using characteristic numbers for the Milky Way Galaxy, 
$M = 6\times 10^{11} M_{\odot}$, $R = 100$ kpc and
$\mu m_p\simeq 1.3$ GeV (which takes the mass of 
the halo to be dominated by completely ionised $He'$,
as suggested by mirror big bang nucleosynthesis) 
we find: $kT_{virial} \approx 100$ eV.
The assumption of complete ionisation is justified 
because
the temperature is greater than the ionisation 
energy of $He'$ (which is about $55$ eV for the second electron).

Given that $He'$ is fully ionised, the electron number density
in the proto-galactic nebula is
\begin{eqnarray}
n_{e'} &=& {3M \over 4\pi R^3}
{2 \over 3\mu m_p}
\nonumber \\
&=& 3\times 10^{-3} \left( {M \over 6\times 10^{11} M_{\odot}}
\right) \left( {100\ {\rm kpc} \over R}\right)^3
\left( {1.3 \ {\rm GeV} \over \mu m_p}\right)
\ {\rm cm^{-3}}.
\label{typ}
\end{eqnarray}
Interactions of mirror electrons with mirror ions will produce
mirror photons via 
several processes, including bremsstrahlung, mirror
electron capture, and so on. The halo is expected to be optically thin to
such mirror photons since their mean scattering length,
\begin{eqnarray}
\ell & = & {1 \over n_{e'} \sigma_T}
\nonumber \\
& \approx & \left[ {3\times 10^{-3}\ {\rm cm^{-3}} \over n_{e'}}\right]
2\times 10^5 \ {\rm kpc},
\end{eqnarray}
is much larger than a galactic radius. Here
$\sigma_T$ is the Thomson cross section: 
$\sigma_T \simeq 6.65\times 10^{-25}\ {\rm cm^2}$.
Thus, any mirror photons produced should escape 
the galaxy, thereby cooling it.
The cooling rate will be proportional to the product
of the mirror electron and mirror ion number densities.
Since the gas is highly ionised, the mirror ion number
density is roughly half the mirror electron
number density. Thus, 
the cooling rate per unit volume, $\Gamma_{cool}$, for dissipative
processes can be considered as proportional to $n_{e'}^2$:
\begin{eqnarray}
\Gamma_{cool} = n_{e'}^2 \Lambda \ .
\end{eqnarray}
The quantity $\Lambda$ contains the details of the cross section,
temperature, and so on.
For a temperature of $T_{virial} \sim 100$ eV, $\Lambda \sim 10^{-23}\
{\rm erg \ cm^3\ s^{-1}}$\cite{bin}.

At the virial temperature, the energy per unit volume is
of order $n_{e'} {3 \over 2} kT_{virial}$. It follows that
the time scale for which the radiative cooling would
remove all the energy from the gas is
\begin{eqnarray}
t_{cool} &=& {3 \over 2} {kT_{virial} n_{e'} \over \Gamma_{cool} }
\nonumber \\
 &=& {3 \over 2}{k T_{virial} \over n_{e'} \Lambda}.
\end{eqnarray}
For $n_{e'} \sim 3\times 10^{-3}\ {\rm cm^{-3}}$, 
$t_{cool} \sim 3\times 10^8$ years.
This suggests that a halo composed predominantly of a 
gas of mirror ions and mirror electrons would be 
expected to dissipate energy
too quickly to long endure. We shall call this the
{\it radiative cooling problem.}

\section{Estimate of the halo mirror photon luminosity}

The radiative cooling problem would be solved if there was a heating
mechanism, so that the energy lost due to radiative cooling could
be replaced (possible heating mechanisms will
be discussed later on). 
Assuming for now that this does indeed occur, then the collapse
of the gas is halted and hydrostatic equilibrium holds good.

Taking a spherical dark matter halo, the condition
of hydrostatic equilibrium gives
\begin{eqnarray}
{dP(r) \over dr} = -\rho(r) g(r)
\label{pr}
\end{eqnarray}
where $P(r)$ is the pressure, $\rho(r)$ the mass density and $g(r)$ the
local acceleration, at radius $r$.
For a dilute gas, the pressure
is related to the mass density via $P = \rho kT/(\mu m_p)$,
where $\mu m_p$ is the average mass of the particles in the gas,
$m_p$ being the proton mass. 
Taking the usual case of an isothermal halo, $T$
does not depend on $r$.
The local acceleration can also be simply expressed in terms of the
mass density via
\begin{eqnarray}
g(r) = {4\pi G \over r^2}\int^r_{0}\, \rho(r')\, r'^2\, dr' \ ,
\label{fin}
\end{eqnarray}
where $G$ is Newton's constant.
Equations (\ref{pr}) and (\ref{fin}) can now be solved for $\rho$ to give 
\begin{eqnarray}
\rho = {\lambda \over r^2}, 
\end{eqnarray}
where $\lambda$ is a constant that satisfies
\begin{eqnarray}
k T = 2\pi G\, \lambda\, \mu m_p.
\end{eqnarray}
The rotational velocity at radius $r$ is given by
\begin{eqnarray}
v^2_{rot} &=& {4\pi G \over r}\int^{r}_0\, \rho\, r'^2\, dr'
\nonumber \\
&= & 4\pi G\, \lambda,
\label{flat}
\end{eqnarray}
which is a constant. This is just the usual result that a
$\rho = \lambda/r^2$ behaviour of a spherically-symmetric, isothermal,
self-gravitating gas in hydrostatic equilibrium gives
a flat rotation curve.

Using Eq.(\ref{flat}) to write $\lambda$ in terms of the
rotational velocity, we see that
$\rho$ can be reparameterised as
\begin{eqnarray}
\rho(r) &=& {v^2_{rot}
\over 4\pi G}{1 \over r^2} 
\nonumber \\
&\approx  & 0.3 \left( {v_{rot} \over
220 \ {\rm km/s}}\right)^2 \left({10 \ {\rm kpc} \over
r}\right)^2 \ {\rm \frac{GeV}{c^2}\ cm^{-3}}.
\end{eqnarray}
Note that $n_{e'} = 2n_{He'} \simeq 2\rho/m_{He'}$ 
(for a $He'$ mass dominated halo),
which implies 
\begin{eqnarray}
n_{e'} \approx 10^{-1} \ \left( {10 \ {\rm kpc} \over r}\right)^2 
\ {\rm cm^{-3}},
\end{eqnarray}
having set $v_{rot} \approx 220$ km/s.

Since $n_{e'} \propto 1/r^2$,
the total halo luminosity, 
\begin{eqnarray}
L_{halo} = 4\pi\Lambda\, \int_{r_{min}}^{\infty}\, n_{e'}^2\, r^2\, dr,
\end{eqnarray}
is divergent as $r_{min} \to 0$. 
However, the inner region of the galaxy
should contain a high density of mirror dust, mirror stars, 
mirror supernovas,
blackholes, and so on, which complicates the situation considerably. 
For example, it is possible that the temperature
increases towards the galactic centre due to the presence of:
a) heat sources such as supernovas (see
later discussion) and b) mirror dust particles, which can
potentially make the inner region optically thick to
mirror radiation.
If this were the case, then
the isothermal approximation would be invalid and then
the mass density need not continue to increase as $1/r^2$ as
$r \to 0$. This would also be
consistent with observations of
rotation curves in spiral galaxies\cite{salucci1}.
These observations suggest that the mass density
is roughly constant in the inner and central regions of 
spiral galaxies, as if the halo were ``heated up'' (in
the vernacular of Ref.\cite{salucci2}) in the inner
region.

Thus, we introduce a phenomenological cutoff, $R_1$, and
consider only the energy produced for $r > R_1$.
In this case, the energy radiated from the halo is
roughly
\begin{eqnarray}
L_{halo} &=& 4\pi\Lambda\, \int_{R_1}^{100 {\rm kpc}}\, n_{e'}^2\, r^2 dr
\nonumber \\
&\sim &
\left({3 \ {\rm kpc} \over R_1}
\right)
10^{44}
\ \ {\rm erg/s}.
\end{eqnarray}
The above calculation assumes that the halo contains only
a gas component. As discussed earlier,
a significant component of the halo will be in the form of compact
mirror objects. Furthermore, they
can potentially dominate the mass in the inner regions
of the galaxy -- which would alleviate the cooling problem 
to some extent.
Still, a heat source of at least $10^{43}$ erg/s
seems to be required to replace the energy lost due to
radiative cooling.

\section{Galactic heating sources}

We now examine possible heating sources that could compensate for
the energy lost due to radiative cooling.
Perhaps the most obvious energy sources are supernova explosions,
both ordinary and mirror types.
Mirror supernovas can supply the mirror halo with around $10^{51}$ erg per
explosion (this is the kinetic energy of the outer layers
ejected into the interstellar medium).
To account for the radiative energy loss
would require a galactic mirror supernova explosion rate of
around one per year.  
This rate is about two orders of magnitude larger than
the rate of ordinary supernovas in our galaxy.
However, the ordinary and mirror sectors have different chemical 
compositions, abundances and
distributions. There is no macroscopic mirror symmetry. 
It follows that there is no reason for the rates of
ordinary and mirror supernovas to be the same.

Another interesting possibility is that ordinary supernovas
could supply this missing energy.  While the kinetic energy of 
the ejected outer layers of a supernova is of order $10^{51}$ erg, a
supernova has a total energy output $E_{SN}$ of about $3\times 10^{53}$ erg. 
In standard theory, this energy is released into neutrinos. 
However, a substantial portion of this energy can be converted into
mirror electrons, mirror positrons and mirror photons if photon--mirror-photon 
kinetic mixing 
exists.\footnote{Likewise, a mirror
supernova would be a source of ordinary electrons, positrons and
photons and may be related to
observations of Gamma Ray bursts and galactic 511 keV photons\cite{fs}.} 
This particle interaction has the explicit form 
\begin{eqnarray}
{\cal{L}} = {\epsilon \over 2} F^{\mu \nu} F'_{\mu \nu} 
\label{km}
\end{eqnarray}
where $F^{\mu \nu}$ and $F'^{\mu \nu}$ are the
electromagnetic field strength tensors for ordinary and mirror
electromagnetism. One effect of this interaction is to cause mirror
charged particles (such as the mirror electron and mirror proton)
to couple to ordinary photons with effective electric charge
of $\epsilon e$\cite{flv,hol,s}.
Thus the parameter $\epsilon$ determines the 
strength of this interaction,
with $\epsilon \sim 5\times 10^{-9}$ suggested from a fit\cite{footdama} 
to the DAMA/NaI annual
modulation signal\cite{dama}.
Also, for $\epsilon \sim 10^{-9}$ mirror particle emission from
an ordinary supernova is comparable to neutrino emission\cite{raf,fs}.
In other words,
a significant fraction, $f'$, of an ordinary supernova's
{\it total} energy can be released into mirror electrons, positrons and mirror
photons for epsilon values near what is preferred by DAMA/NaI. If $f' \sim 0.1$,
then the energy converted into mirror particle production is actually 
an order of magnitude larger than that converted into ordinary particle 
kinetic energy!
In some circumstances, shocks will develop which will
accelerate the plasma to form a mirror gamma ray burst.
This may require special circumstances (such as right the
amount of ordinary baryons). 

Whatever the intangibles, a
significant proportion of a supernova's total
energy may be released into $e'^{\pm}$ and $\gamma'$.
If so, the mirror electrons and mirror positrons 
would not escape 
out of the galaxy because they would plausibly 
be confined by the mirror magnetic field and
so could give up most of their energy into heating
the mirror particles in the halo. Furthermore,
mirror photons could be absorbed by heavy mirror elements in the halo
if their energies were in the keV range.
The point is that elements heavier than about mirror carbon, $C'$, can
retain their K-shell mirror electrons, since the
binding energies are greater than the temperature of the
particles in the halo.
The cross section for photoionisation of K-shell
mirror electrons (for atomic number Z) is\cite{bj}
\begin{eqnarray}
\sigma &=& {16 \sqrt{2} \pi \over 3} \alpha^8 Z^5 \left( {m_e c^2 \over
E_{\gamma'}}\right)^{7/2}\ a_0^2
\nonumber \\
&\simeq & 5\times 10^{-19} \left( {Z \over 8}\right)^5 \left( {{\rm keV} \over
E_{\gamma'}
}\right)^{7/2} \ {\rm cm^2},
\end{eqnarray}
where $a_0$ is the Bohr radius,
giving a mean free path of
\begin{eqnarray}
\ell \approx 7 \left( {8 \over Z}\right)^5 \left( {E_{\gamma'}  
\over {\rm keV}}\right)^{7/2} \left( {10^{-4} \ {\rm cm^{-3}}\over n_{A'}}\right)\
{\rm kpc},
\end{eqnarray}
where $n_{A'}$ is the number density of heavy elments
($M_{A'}\stackrel{>}{\sim} M_{C'}$).
This is one plausible way that mirror particles, $e'^{\pm}$ and $\gamma'$, produced
in ordinary supernova explosions could potentially
be absorbed
in the halo -- providing a significant heating source.
The
amount of energy going into the halo from ordinary
supernova explosions is roughly\cite{fs}
\begin{eqnarray}
E_{in} &=& f' E_{SN} \Gamma_{SN}
\nonumber \\
&=& \left( {f' \over 0.1}\right) \left( {E_{SN} \over 3\times 10^{53} \
{\rm erg}}
\right)\left( {\Gamma_{SN} \over 0.01 \ {\rm yr^{-1}}}\right)\ 10^{43}\ {\rm erg/s},
\end{eqnarray}
where $\Gamma_{SN}$ is the galactic supernova rate.
Evidently, ordinary supernovas can potentially supply about
the right amount of energy to replace the energy lost in
radiative cooling, if ordinary supernovas occur at a rate
of order once per hundred years and about $10\%$ of a
supernova's energy is converted into mirror electrons, positrons and 
photons.\footnote{One can speculate that the apparent coincidence between
the energy loss rate and the supernova energy injection rate might arise
as the steady state limit of dynamical evolution of the ordinary-plus-mirror
protogalactic nebula into an actual galaxy.}

\section{Conclusion}

We have examined an important problem facing mirror dark
matter: because mirror dark matter is dissipative, spheroidal halos
around spiral galaxies can cool and potentially collapse on
a time scale much shorter than the age of the galaxy.
We estimated the total halo luminosity to be at least $10^{43}$ erg/s.
In the absence of any significant heat source, the time scale of
the collapse would be around 300 Myr.

However there are potentially significant heat sources. In particular,
both ordinary and mirror supernovas are candidates. Mirror
supernovas can supply the energy if they occur at a rate of around one
per year. Alternatively, ordinary supernovas can do the job if
there exists photon--mirror-photon kinetic mixing, with
$\epsilon \sim 10^{-9}$, roughly consistent with
the value suggested by the DAMA experiment. 
The effect of this interaction is to modify the dynamics of
supernova explosions allowing for a significant fraction of
the total energy to be released into
$e'^{\pm}$ and $\gamma'$. The energy of these particles can
be absorbed by the halo, and can potentially
supply the required energy.
Presumably, there needs to be a significant asymmetry in the 
heating rates during the evolution of the galaxy, to explain why
the ordinary matter has collapsed onto the disk and
the mirror matter has not. But this is
possible because of the lack of any macroscopic mirror symmetry.

\acknowledgments{RRV thanks Maxim Khlopov for interesting discussions,
and Paolo Lipari and Maxim Khlopov for fine hospitality at
the Universit\'{a} di Roma ``La Sapienza'' where the discussions took
place.  We would also like
to thank Stuart Wyithe for valuable comments on
a draft of this article.
This work was supported by the Australian Research
Council.}


\begin{thebibliography}{999}

\bibitem{flv}
R. Foot, H Lew and R. R. Volkas, Phys. Lett.
B272, 67 (1991). The idea that exact unbroken parity symmetry can
be respected by the fundamental interactions if
a set of mirror particles exist was earlier
discussed, prior to the advent of the standard model
of particle physics, in
T. D. Lee and C. N. Yang, Phys. Rev.
104, 256 (1956);
I. Kobzarev {\it et al}.,
Sov. J. Nucl. Phys. 3, 837 (1966); M. Pavsic, Int. J. Theor. Phys. 9,
229 (1974), with application to
dark matter first suggested by
S. I. Blinnikov and M. Yu. Khlopov, Sov. J. Nucl. Phys.
36, 472 (1982); Sov. Astron. 27, 371 (1983).

\bibitem{comelli}
Z. Berezhiani, D. Comelli and F. L. Villante, Phys. Lett. B503, 362
(2001) [hep-ph/0008105]; Z. Berezhiani, P. Ciarcelluti, D. Comelli
and F. L. Villante, astro-ph/0312605.

\bibitem{rays}
A. Yu. Ignatiev and R. R. Volkas, Phys. Rev. D68, 023518 (2003)
[hep-ph/0304260].



\bibitem{cdms2}
D. S. Akerib {\it et al.} (CDMS Collaboration),
astro-ph/0405033.

\bibitem{ed}
A. Benoit {\it et al.} (Edelweiss Collaboration),
Phys. Lett. B545, 43 (2002) [astro-ph/0206271].

\bibitem{dama}
R. Bernabei et al. (DAMA Collaboration), 
Phys. Lett. B480, 23 (2000);
Riv. Nuovo Cimento. 26, 1 (2003) [astro-ph/0307403].

\bibitem{footdama}
R. Foot, Phys. Rev. D69, 036001 (2004) [hep-ph/0308254];
astro-ph/0403043; R. Foot, Mod. Phys. Lett. A19, 1841 
(2004) [astro-ph/0405362].

\bibitem{review}
See e.g.
A. Yu. Ignatiev and R. R. Volkas, hep-ph/0306120;
R. Foot, to appear.

\bibitem{macho}
C. Alcock {\it et al.} (MACHO Collaboration), Ap. J. 542, 281(2000)
[astro-ph/0001272].

\bibitem{new}
R. Uglesich {\it et al}., astro-ph/0403248.

\bibitem{machoexp}
Z. Silagadze, Phys. Atom. Nucl. 60, 272 (1997) [hep-ph/9503481];
S. Blinnikov, astro-ph/9801015; R. Foot, Phys. Lett. B452, 83 (1999)
[astro-ph/9902065].

\bibitem{old}
R. Foot and R. R. Volkas,
Astropart. Phys. 7, 283 (1997) [hep-ph/9612245].


\bibitem{wmap}
D. N. Spergel {\it et al},
(WMAP Collaboration) Astrophys. J. Suppl. 148, 175 (2003)
[astro-ph/0302209]
and references therein for earlier work.

\bibitem{inflation}
E. W. Kolb, D. Seckel and M. S. Turner, Nature 314, 415 (1985);
H. M. Hodges, Phys. Rev. D47, 456 (1993); Z. G. Berezhiani, A. D. Dolgov
and R. N. Mohapatra, Phys. Lett. B375, 26 (1996) [hep-ph/9511221];
V. Berezinsky and A. Vilenkin, Phys. Rev. D62, 083512 (2000)
[hep-ph/9908257].

\bibitem{fv}
R. Foot and R. R. Volkas, Phys. Rev. D68, 021304 (2003)
[hep-ph/0304261]; Phys. Rev. D69, 123510 (2004) [hep-ph/0402267]
see also:
L. Bento and Z. Berezhiani, Phys. Rev. Lett. 87, 231304 (2001) [hep-ph/0107281];
hep-ph/0111116.

\bibitem{astro}
B. W. Carroll and D. A. Ostlie,
{\it An Introduction to Modern Astrophysics}, 
Addison-Wesley, 1995.

\bibitem{bin}
J. Binney and S. Tremaine, 
{\it Galactic Dynamics}, Princeton University Press,
Princeton, NJ, 1987.

\bibitem{salucci1}
See e.g.
P. Salucci and A. Borriello,
MNRAS, 323, 285
(2001) [astro-ph/0001082];
E. D'Onghia {\it et al}.,
astro-ph/0107423.

\bibitem{salucci2}
P. Salucci and A. Borriello,
astro-ph/0106251.

\bibitem{hol}
B. Holdom, Phys. Lett. B166, 196 (1986).

\bibitem{s}
R. Foot, A. Yu. Ignatiev and R. R. Volkas,
Phys. Lett. B503, 355 (2001) [astro-ph/0011156].
										

\bibitem{raf}
G. Raffelt, {\it Stars as Laboratories for fundamental physics},
Chicago University Press (1995).
                                                                                

\bibitem{fs}
R. Foot and Z. K. Silagadze, 
astro-ph/0404515.

\bibitem{bj}
See e.g.
B. H. Bransden and C. J. Joachain, {\it
Physics of atoms and molecules}, Longman, p.193 (1983).


\end{thebibliography}
\end{document}